\documentclass[
 reprint,amsmath,amssymb,aps,showpacs,superscriptaddress
]{revtex4-1}

\usepackage{graphicx}
\usepackage{dcolumn}
\usepackage{bm}
\usepackage{here}
\usepackage{multirow}
\usepackage{color}

\begin{document}

\title{Role of breakup processes in deuteron-induced spallation reactions at 100--200 MeV/nucleon}

\author{Shinsuke Nakayama}
\email{nakayama.shinsuke@jaea.go.jp}
\affiliation{Nuclear Data Center, Japan Atomic Energy Agency, Ibaraki 319-1195, Japan }
\author{Naoya Furutachi}
\affiliation{Nuclear Data Center, Japan Atomic Energy Agency, Ibaraki 319-1195, Japan }
\author{Osamu Iwamoto}
\affiliation{Nuclear Data Center, Japan Atomic Energy Agency, Ibaraki 319-1195, Japan }
\author{Yukinobu Watanabe}
\affiliation{Department of Advanced Energy Engineering Science, Kyushu University, Fukuoka 816-8580, Japan}

\date{\today}

\begin{abstract}
\begin{description}
\item[Background]
Use of deuteron-induced spallation reactions at intermediate energies has recently been proposed for transmutation of several long-lived fission products (LLFPs). 
In the design study of a transmutation system using a deuteron primary beam, accurate cross section data of deuteron-induced reactions on the LLFPs are indispensable. 
Reliable model predictions play an important role in completing the necessary cross section data since currently available experimental data are very limited. 
Under the circumstances, we have been developing a code system dedicated for deuteron-induced reactions, which is called DEURACS.
\item[Purpose]
Aiming to predict the production cross sections of residual nuclei,  
the purpose of the present work is to clarify a role of deuteron breakup processes in deuteron-induced spallation reactions at intermediate energies.
\item[Methods]
Isotopic production cross sections of residual nuclei in the deuteron-induced reactions on $^{93}$Zr and $^{107}$Pd at 100--200 MeV/nucleon are analyzed using DEURACS,
in which the breakup processes are explicitly taken into account.
The calculated cross sections are decomposed into individual components corresponding to the absorption of either neutron or proton in the incident deuteron, or the deuteron itself.
\item[Results]
The calculated cross sections reproduced the experimental data well over a wide mass number range of residual nuclei.
From a component-by-component analysis, it was found that the components of nucleon absorption have the significant contributions to the production of residual nuclei.
\item[Conclusions]
Consideration of the breakup processes is essential to predict the production cross sections of residual nuclei in deuteron-induced reactions.
The framework of DEURACS is applicable to deuteron-induced spallation reactions at intermediate energies. 
\end{description}
\end{abstract}

\maketitle

\section{Introduction}
\label{sec:intro}
Hundreds of nuclear power plants have been operated and have contributed to energy supply in the world.
Among radionuclides produced in a nuclear reactor, those with long half-lives require long-term management beyond the history of civilization 
and hence it is strongly desired to convert them into stable or short-lived ones.
Spallation reaction is one of the candidates especially for transmutation of radioactive nuclides whose neutron capture cross sections are small~\cite{WangCs}.
Quite recently, several experimental studies on spallation reactions have been conducted for transmutation of $^{93}$Zr~\cite{Kawase} and $^{107}$Pd~\cite{WangPd}, 
which are typical long-lived fission products (LLFPs) with the half-lives of $1.6\times10^6$ and $6.5\times10^6$ years, respectively.
In Refs.~\cite{WangPd,Kawase}, proton- and deuteron-induced spallation reactions at 100–-200 MeV/nucleon were measured in inverse kinematics 
and it was suggested that both reactions can be effectively used for transmutation of the LLFPs.
Additionally, as mentioned in Ref.~\cite{Kawase}, the deuteron may be a better candidate than the proton. 
This is because deuteron-induced reactions are expected to emit more secondary neutrons, which can contribute to further transmutation in a target system.

In the design study of a transmutation system using a deuteron primary beam, accurate cross section data of deuteron-induced reactions on LLFPs are indispensable.
Isotopic production cross sections of residual nuclei are especially important to estimate the conversion rates into stable or short-lived nuclei and the production rates of other undesirable long-lived ones.
However, currently available experimental data of deuteron-induced reactions on LLFPs are very limited due to the difficulty of measurements.
In such a case, reliable theoretical model calculations play a key role in completing the necessary cross section data by interpolating or extrapolating experimental ones.

The deuteron is a weakly bound nucleus and easily breaks up by the interaction with a target nucleus. 
By previous experimental~\cite{15MeV,25MeV,56MeV,80MeV,100MeV,Wakasa,Araki} and theoretical~\cite{Lei-Moro1,Lei-Moro2,Potel,Carlson,Lei-Moro3,Potel2,YePRC2009,YePRC2011} studies
including ours~\cite{NakayamaND2013,NakayamaINES4,NakayamaJNST,NakayamaCNR15,NakayamaPRC,NakayamaND2016},
it is well-known that breakup processes play a significant role in the secondary nucleon emission from deuteron-induced reactions.
Similarly, the breakup processes are expected to have some influence on spallation reactions.
Understanding of the influence is essential to predict the production cross sections of residual nuclei.

As for theoretical approach to the production of residual nuclei in deuteron-induced reactions, 
M. Avrigeanu and her co-workers have proposed a method of considering the breakup processes and have achieved great success 
for the reactions with the total kinetic energy below 60 MeV~\cite{AvrAl,AvrCu,AvrAC,AvrNb,AvrFe,AvrRole,AvrNi}.
In their method, the deuteron breakup processes are described by the empirical formula based on the experimental data up to 80 MeV~\cite{AvrBU}.
It is controversial whether their method works well for incident energies of 100--200 MeV/nucleon of interest in the present work, 
which largely exceed the applicable range of the formula.

Under the above situations, we have been developing a code system dedicated for deuteron-induced reactions, which is called deuteron-induced reaction analysis code system (DEURACS)~\cite{NakayamaPRC,NakayamaND2016}. 
DEURACS consists of several calculation codes based on theoretical models to describe respective reaction mechanisms characteristic of deuteron-induced reactions.
The deuteron breakup processes, which are classified into elastic breakup and non-elastic breakup processes, are calculated using the codes based on the continuum-discretized coupled-channels (CDCC) theory~\cite{CDCC} and the Glauber model~\cite{YePRC2009,YePRC2011}, respectively. 
In addition to that, the pre-equilibrium and the compound nucleus processes are calculated using the CCONE code~\cite{CCONEJNST2007,CCONENDS2016}.

One of the main purposes to develop DEURACS was to contribute to the design study of intensive neutron sources using deuteron accelerators.
Hence, DEURACS had been mainly applied to analysis of double-differential cross sections of the $(d,xn)$ and $(d,xp)$ reactions 
and the calculated results were in good agreement with the experimental data up to 50 MeV/nucleon especially at forward angles~\cite{NakayamaND2013,NakayamaINES4,NakayamaJNST,NakayamaCNR15,NakayamaPRC,NakayamaND2016}.
From these analyses it was concluded that the models implemented in DEURACS to calculate the breakup processes are valid in the energy range 
since the nucleon emission at forward angles has a large sensitivity to treatment of the breakup processes.

The CDCC method and the Glauber model have been widely used in analyses at incident energies above 200 MeV/nucleon~\cite{CDCC,Glauber1,Glauber2}.
Therefore, it is expected that the framework of DEURACS for the breakup processes is applicable to the incident energy range of 100--200 MeV/nucleon.
It is of interest to apply DEURACS to deuteron-induced spallation reactions at this energy range.
In the present work, isotopic production cross sections of residual nuclei in the deuteron-induced reactions on $^{93}\mathrm{Zr}$ and $^{107}\mathrm{Pd}$ at 100--200 MeV/nucleon are calculated with DEURACS.
Through comparisons of the calculated cross sections with the measured data, a role of deuteron breakup processes is discussed.

This article is organized as follows.
Section \ref{sec:model} describes a method of calculating production cross sections of residual nuclei and theoretical reaction models implemented in DEURACS. 
Input parameters used in the model calculations are also explained here. 
In Sec.~\ref{sec:result}, calculation results are compared with experimental data and discussions are made. 
Finally, a summary and conclusions are given in Sec.~\ref{sec:conclusion}.

\section{Models and Methods}
\label{sec:model}
\subsection{Calculation method for production cross section of residual nuclei}
\label{sec:models}
In deuteron-induced reactions, three types of composite nuclei can be formed by the absorption of either neutron or proton in the incident deuteron, or the deuteron itself.
In DEURACS, a calculation taking these effects into account are performed by combining the Glauber model and the CCONE code.
The former is used to calculate the direct processes leading to formation of a highly excited nucleus, namely the non-elastic breakup and the deuteron absorption. 
Note that the elastic breakup does not contribute to formation of an excited nucleus since the target nucleus remains in its ground state.  
Same as our preceding works~\cite{NakayamaPRC, NakayamaND2016}, a noneikonal approach is incorporated into the Glauber model as in Ref.~\cite{YePRC2011}. 
In this approach, the eikonal $S$ matrices used in the Glauber model are replaced by the quantum $S$ matrices given by the optical model calculations.
On the other hand, pre-equilibrium and compound nucleus processes are calculated by the exciton model and the Hauser–-Feshbach statistical model implemented in the CCONE code, respectively.
The formulation of two-component exciton model by Kalbach~\cite{Kalbach1985, Kalbach1986} and the Hauser–-Feshbach model with the width fluctuation correction~\cite{Width-Fluc} are adopted in CCONE.

In the framework of DEURACS, the production cross section of a residual nucleus $B$ from $A(d,x)$ reaction is expressed by the incoherent summation of the following three components:
\begin{equation}
	\label{equ:sum}
	\sigma^B_{(d,x)} = \sigma^B_{d}+\sigma^B_{p}+\sigma^B_{n},
\end{equation}
where $\sigma^B_{d}$, $\sigma^B_{p}$, and $\sigma^B_{n}$ correspond to the contributions from the absorption of deuteron, proton, and neutron, respectively.
Each component of Eq.\ (\ref{equ:sum}) is given by
\begin{equation}
\label{equ:component}
\sigma^B_{i} = \sigma_{(d,rea)} R_{i} P^B_{i} \ \ \ \ {\rm with} \ i=d,p,n,
\end{equation}
where $\sigma_{(d,rea)}$ is the deuteron total reaction cross section.
$R_{i}$ is the formation fractions of composite nucleus by the absorption of particle $i$.
$P^B_{i}$ is the probability that $B$ is finally produced when $i$ is absorbed.

Each $R_{i}$ is given as follows:
\begin{eqnarray}
\label{equ:Rd}
&R_{d}& = \frac{\sigma_{d-\mathrm{ABS}}}{\sigma_{(d,rea)}},\\
\label{equ:Rp}
&R_{p}& = \frac{\sigma_{p-\mathrm{NEB}}}{\sigma_{(d,rea)}},\\
\label{equ:Rn}
&R_{n}& = \frac{\sigma_{n-\mathrm{NEB}}}{\sigma_{(d,rea)}},
\end{eqnarray}
where ${\sigma_{(d,rea)}}$ is the deuteron total reaction cross section obtained by the Glauber model,
and $\sigma_{d-\mathrm{ABS}}$, $\sigma_{p-\mathrm{NEB}}$, and $\sigma_{n-\mathrm{NEB}}$ denote the cross sections for the deuteron absorption, the non-elastic breakup where proton is absorbed,
and the neutron removal non-elastic breakup where neutron is absorbed calculated with the Glauber model, respectively.

Since the deuteron is absorbed in the target nucleus at the certain incident energy, $P^B_{d}$ is calculated as follows:
\begin{equation}
\label{equ:pbd}
P^B_{d} = \frac{\sigma_{A(d,x)B}}{\sigma_{(d,rea)}},
\end{equation}
where $\sigma_{A(d,x)B}$ is the production cross section of $B$ from deuteron-induced reactions calculated with the CCONE code.
On the other hand, the proton or the neutron absorbed in the target nucleus after the non-elastic breakup has an energy distribution.
Thus, we calculate $P^B_{p}$ and $P^B_{n}$ taking the distribution into account:
\begin{equation}
\label{equ:pbp}
P^B_{p} = \int dE_p f(E_p) \frac{\sigma_{A(p,x)B}(E_p)}{\sigma_{(p,rea)}(E_p)},
\end{equation}
\begin{equation}
\label{equ:pbn}
P^B_{n} = \int dE_n f(E_n) \frac{\sigma_{A(n,x)B}(E_n)}{\sigma_{(n,rea)}(E_n)},
\end{equation}
where $f(E_p)$ and $f(E_n)$ are the normalized energy distributions of the absorbed proton and neutron. 
They are obtained kinematically from the spectra of the neutron and proton emitted by the non-elastic breakup, which are calculated with the Glauber model, respectively.
$\sigma_{A(p,x)B}$ and $\sigma_{A(n,x)B}$ are the production cross section of $B$ from proton- and neutron-induced reactions calculated with the CCONE code,
and $\sigma_{(p,rea)}$ and $\sigma_{(n,rea)}$ are the proton and neutron total reaction cross sections.

\subsection{Input parameters of model calculations}
 \label{sec:parameters}
 The theoretical models described in Sec.~\ref{sec:models} use some input parameters.
First, in the Glauber model and the CDCC method, nucleon optical potentials (OPs) at half the incident deuteron energy are necessary. 
The phenomenological global nucleon OPs of Koning and Delaroche (KD)~\cite{K-DOMP} have been widely used in the incident energy range up to 200 MeV.
Since 200 MeV/nucleon is the upper limit of the incident deuteron energy in the present work, the KD OPs are adopted for both proton and neutron.
Following Ref.~\cite{Neoh}, we have taken into account the $p$-$n$ partial waves up to $\ell_{\mathrm{max}}=6$ in the CDCC method.
Each of them is discretized by the momentum-bin method with an equal increment $\Delta k = 0.1$ $\mathrm{fm}^{-1}$ to a maximum of $k_{\mathrm{max}} = 1.0$ $\mathrm{fm}^{-1}$.

Next, in the CCONE calculation, the phenomenological global OPs of An and Cai (AC)~\cite{AnCaiOMP} and the KD OPs are used for deuteron and nucleon, respectively.
The pre-equilibrium model parameters such as the potential depth $V$ and the effective squared matrix element $M^2$ are taken from the work by Koning and Duijvestijn~\cite{preeq}.
As for level densities, the composite formula of Gilbert and Cameron~\cite{Gilbert} are adopted and the systematics of Mengoni and Nakajima~\cite{Mengoni} are used for the parameterization of the Fermi-gas model.
The default values in CCONE are employed for other input parameters.

Finally, we mention the deuteron total reaction cross section $\sigma_{(d,rea)}$, which is one of the most important quantities in the present calculation.  
The applicable range of the AC OPs is set to be up to about 100 MeV/nucleon~\cite{AnCaiOMP} and it is not clear whether $\sigma_{(d,rea)}$ calculated with the OPs is reliable above the upper limit.
Although there is no global deuteron OP applicable up to 200 MeV/nucleon, more recently Minomo, Washiyama, and Ogata (MWO) have proposed a formula describing $\sigma_{(d,rea)}$ up to 500 MeV/nucleon
based on the calculation results by a microscopic three-body reaction model~\cite{MWO}.
Hence, we perform the CCONE calculation in the framework of DEURACS up to 200 MeV/nucleon using the AC OPs as the deuteron OPs, 
and then re-normalize the calculation results using the $\sigma_{(d,rea)}$ by the MWO formula.
In other words, the $\sigma_{(d,rea)}$ by the formula is used for $\sigma_{(d,rea)}$ in Eq.~(\ref{equ:component}).
We discuss the details of this prescription later in Sec.~\ref{sec:d-rea}.

\section{Results and discussions}
\label{sec:result}
\subsection{$^{93}\mathrm{Zr}, ^{107}\mathrm{Pd}+d$ reactions around 100 MeV/nucleon}
\label{sec:100MeV}
\begin{figure}[h]
	\includegraphics[scale=0.90]{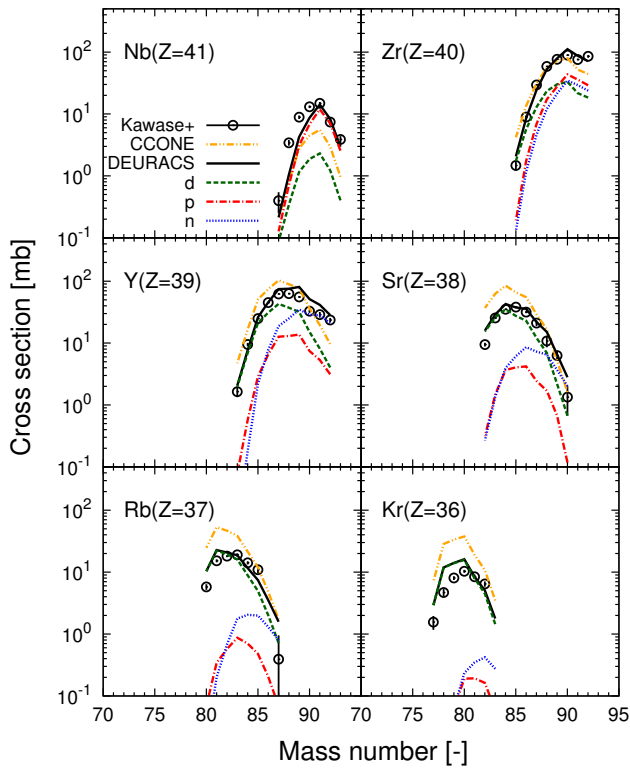}
	\centering
	\caption{\label{fig:zr-105} Calculated and experimental isotopic production cross sections for the $^{93}\mathrm{Zr}+d$ reaction at 105 MeV/nucleon.
								The dashed, dash-dotted, and dotted curves represent the components from the absorption of deuteron, proton, and neutron, respectively. 
								The solid curves are the sum of them. Calculation results of the CCONE code are shown by the dash-dot-dotted curves. 
								The experimental data are taken from Ref.~\cite{Kawase}.}
\end{figure}

Figure~\ref{fig:zr-105} shows the calculated and experimental isotopic production cross sections for the $^{93}\mathrm{Zr}+d$ reaction at 105 MeV/nucleon.
The experimental data are taken from Ref.~\cite{Kawase}.
The calculated cross sections are decomposed into the three components as expressed in Eq.~(\ref{equ:sum}) and each component is also shown in the figure. 
For comparison, we also present the calculation results using only the CCONE code, in which the deuteron breakup processes are not taken into account.
This means that $\sigma_{(d,rea)}$ is simply regarded as the cross section for the deuteron absorption since the direct inelastic scattering is ignored in the present study.
Note that the normalization using the $\sigma_{(d,rea)}$ by the MWO formula is applied also to the CCONE calculation. 

Obvious differences are seen between the calculations of DEURACS and CCONE in the wide range of mass number $A$. 
As an overall trend, CCONE underestimates the production of isotopes with $A>90$ and conversely overestimates the production of those with $A<90$.
On the other hand, DEURACS reproduces the magnitudes and distributions of the experimental data excellently in the wide mass number range of residual nuclei. 
These differences are due to the explicit treatment of the deuteron breakup processes in DEURACS, as discussed in detail below.

As for the production of isotopes with $A>90$, the components of the proton and neutron absorption after the non-elastic breakup make large contributions.
The magnitudes of the two components are comparable to or exceed those of the deuteron absorption components.
The three components have various magnitudes and distributions different from each other and the sum of them reproduces the experimental data well.
These results demonstrate that the deuteron breakup processes play a significant role in spallation reactions.

\begin{figure}[h]
	\includegraphics[scale=0.80]{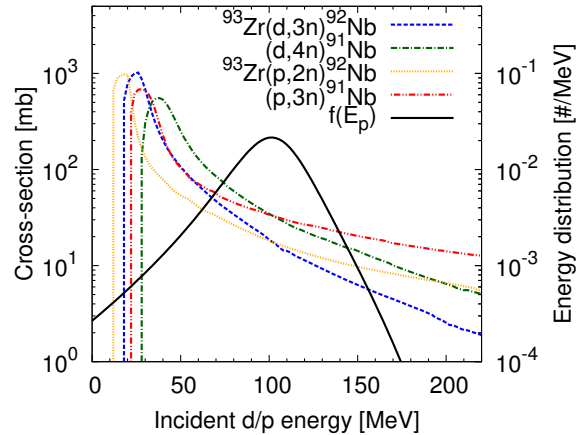}
	\centering
	\caption{\label{fig:nb-xs} Incident energy dependence of the $^{93}\mathrm{Zr}(d,xn)^{95-x}\mathrm{Nb}$ and $^{93}\mathrm{Zr}(p,xn)^{94-x}\mathrm{Nb}$ reactions.
		Solid curve represents $f(E_p)$ at the incident deuteron energy of 105 MeV/nucleon (see Eq.~(\ref{equ:pbp})).
		Note that the horizontal axis denotes the total kinetic energy in MeV.}
\end{figure}

In addition, we can notice that the Nb isotopes are dominantly produced from the proton absorption.
This is because the Nb isotopes are produced only from the $^{93}\mathrm{Zr}(d,xn)^{95-x}\mathrm{Nb}$ and $^{93}\mathrm{Zr}(p,xn)^{94-x}\mathrm{Nb}$ reactions.
Figure~\ref{fig:nb-xs} shows some examples of the calculated incident energy dependence of the Nb isotope production. 
As illustrated in the figure, the production cross sections of the Nb isotopes sharply decrease as the incident energy increases since they are not produced from other reaction channels.
In DEURACS, the proton absorption components are obtained by the convolution using $f(E_p)$ as described in Eq.~(\ref{equ:pbp}). 
Taking the case of the incident energy of 105 MeV/nucleon as an example, this means that the proton is likely to be absorbed around 100 MeV 
while the deuteron is absorbed at the total kinetic energy of 210 MeV. 
Thus, the proton absorption components get much larger than the deuteron absorption ones in the case of the Nb isotopes in Fig.~\ref{fig:zr-105}.

In contrast, the components from the deuteron absorption are found to be predominant in the production of isotopes with $A<90$.
As discussed above, the nucleus formed by the deuteron absorption is likely to induce more highly excited states than those formed by the nucleon absorption.
Hence, residual nuclei with emission of many particles are preferentially produced from the deuteron absorption. 
While CCONE regards $\sigma_{(d,rea)}$ as the deuteron absorption cross sections,
DEURACS divides $\sigma_{(d,rea)}$ into the cross sections corresponding to the elastic breakup, the non-elastic breakup, and the deuteron absorption.
This leads to the reduction of the production cross sections of relatively light isotopes and 
results in a good agreement with the experimental data as illustrated in Fig.~\ref{fig:zr-105}.
Also from these results, it can be said that the consideration of the breakup processes is essentially important in deuteron-induced spallation reactions.

\begin{figure}[h]
	\includegraphics[scale=0.90]{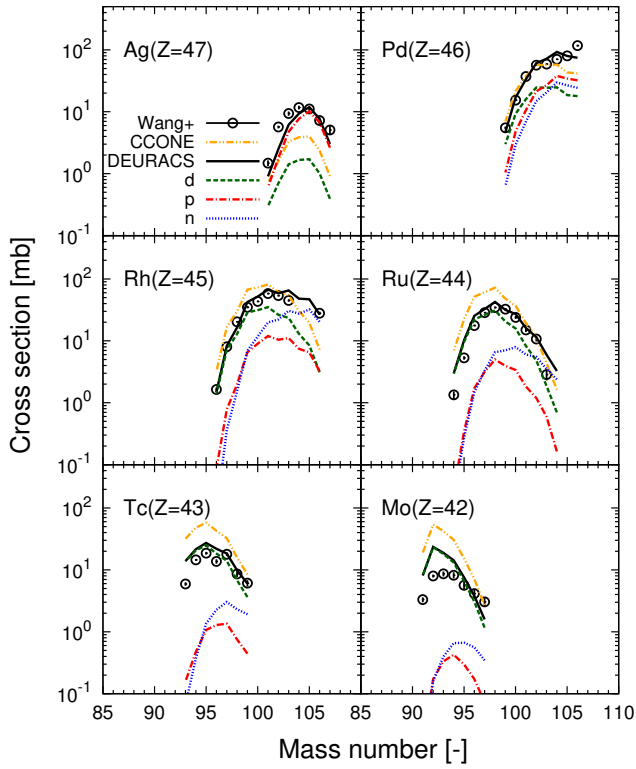}
	\centering
	\caption{\label{fig:pd-118} Same as Fig.~\ref{fig:zr-105} but for the $^{107}\mathrm{Pd}+d$ reaction at 118 MeV/nucleon. 
		The experimental data are taken from Ref.~\cite{WangPd}.}
\end{figure}

Next, to investigate the applicability of DEURACS to another target nucleus, the $^{107}\mathrm{Pd}+d$ reaction at 118 MeV/nucleon is analyzed.
The results are presented in Fig.~\ref{fig:pd-118}.
The experimental data are taken from Ref.~\cite{WangPd}.
As shown in the figure, the DEURACS calculation is in overall good agreement with the experimental data.
Notable features are similar to the case of $^{93}\mathrm{Zr}$ target and as follows:
(i) components of the nucleon absorption have large contributions to production of residual nuclei near the target nucleus.
(ii) isotopes with atomic number $Z$ one larger than the target nucleus, or Nb isotopes for $^{93}\mathrm{Zr}$ target and Ag isotopes for $^{107}\mathrm{Pd}$ target, are dominantly produced from the proton absorption.
(iii) reduction of the deuteron absorption leads to the accurate prediction of cross sections for light residual nuclei.

\subsection{$^{107}\mathrm{Pd}+d$ reaction around 200 MeV/nucleon}
\label{sec:200MeV}
\begin{figure}[h]
	\includegraphics[scale=0.90]{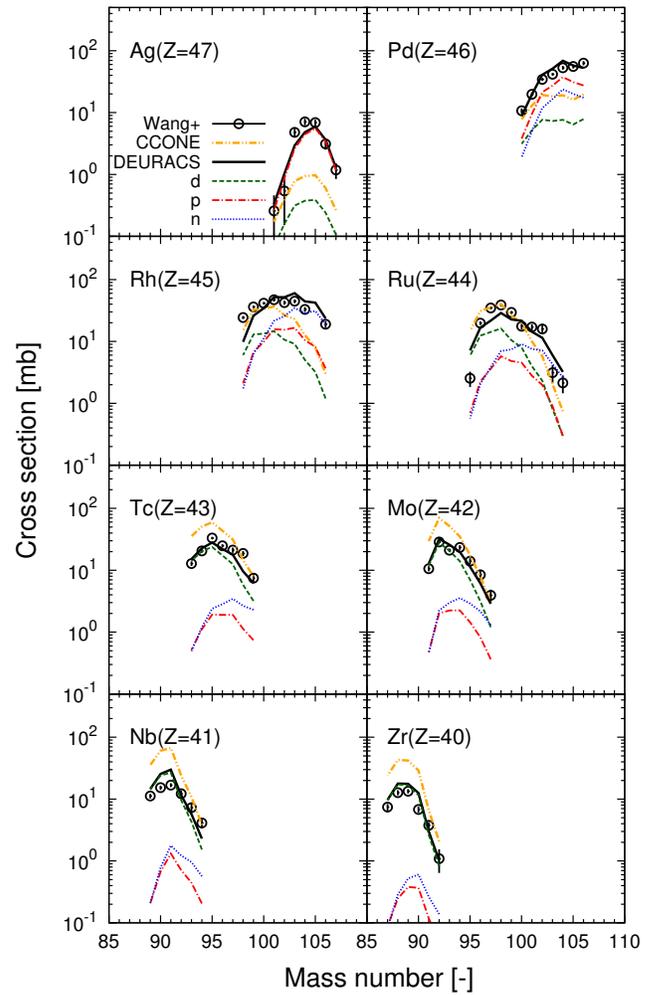}
	\centering
	\caption{\label{fig:pd-196} Same as Fig.~\ref{fig:zr-105} but for the $^{107}\mathrm{Pd}+d$ reaction at 196 MeV/nucleon. 
		The experimental data are taken from Ref.~\cite{WangPd}.
	}
\end{figure}

In order to understand more deeply the relations among the three components, we have analyzed the $^{107}\mathrm{Pd}+d$ reaction at 196 MeV/nucleon.
The results are presented in Fig.~\ref{fig:pd-196}. 
From the figure, it can be confirmed that DEURACS is applicable to the $^{107}\mathrm{Pd}+d$ reaction around 200 MeV/nucleon to the same extent as that around 100 MeV/nucleon. 

\begin{figure}[h]
	\includegraphics[scale=0.90]{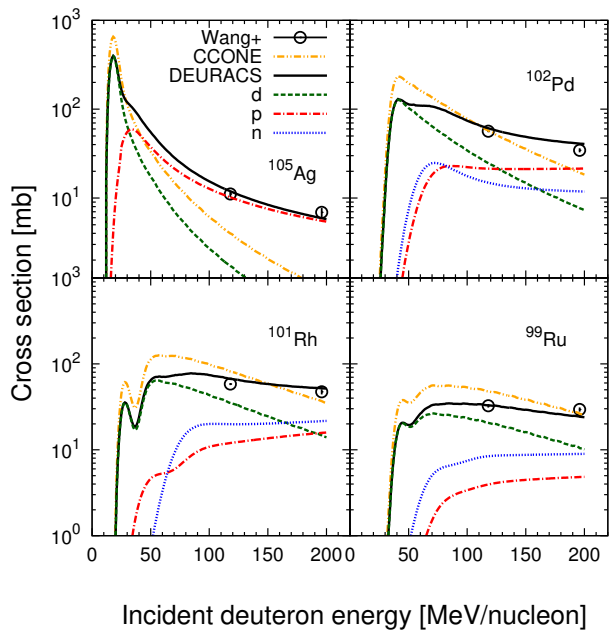}
	\centering
	\caption{\label{fig:pd-ef} Incident energy dependence of the production of residual nuclei for the $^{107}\mathrm{Pd}+d$ reaction.
		The meaning of each curve is same as Fig.~\ref{fig:zr-105}. 
		The experimental data are taken from Ref.~\cite{WangPd}.
	}
\end{figure}

\begin{figure}[h]
	\includegraphics[scale=0.80]{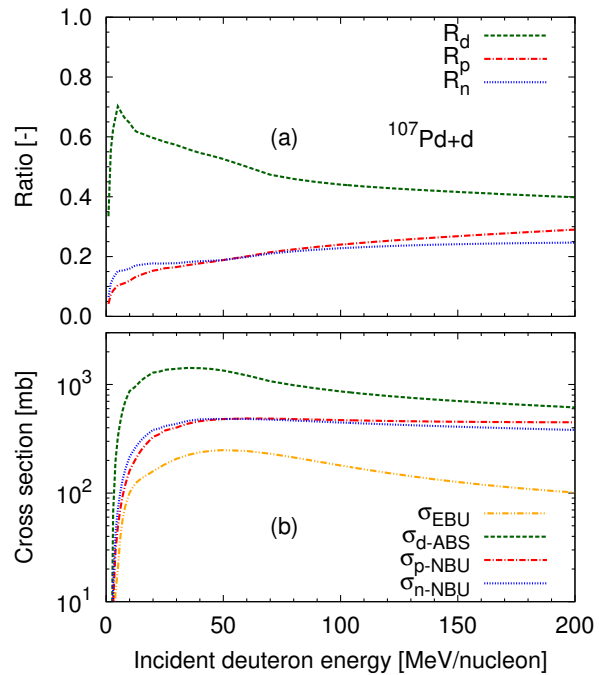}
	\centering
	\caption{\label{fig:Ri-BU} 	
		(a) Incident energy dependence of the formation fractions of composite nucleus for the $^{107}\mathrm{Pd}+d$ reaction.
		The dashed, dash-dotted, and dotted curves represent $R_d$, $R_p$, and $R_n$ in Eq.~(\ref{equ:component}), respectively.
		(b) Incident energy dependence of the direct reaction cross sections for the deuteron interaction with $^{107}\mathrm{Pd}$.
		The cross sections for the elastic breakup are shown by the dash-dot-dotted curves. 
		The dashed, dash-dotted, and dotted curves represent the cross sections for the deuteron absorption, the non-elastic breakup where proton is absorbed,
		and the non-elastic breakup where neutron is absorbed, respectively.
	}
\end{figure}

Based on the results obtained from Figs.~\ref{fig:pd-118} and \ref{fig:pd-196}, we discuss the incident energy dependence of the relations among the three components.
The incident energy dependence of the production cross sections of residual nuclei decomposed into each component is presented in Fig.~\ref{fig:pd-ef}.
The figure shows some examples for the residual nuclei with relatively large contributions from the components of the proton or neutron absorption.
In addition to Fig.~\ref{fig:pd-ef}, the incident energy dependence of each $R_i$ in Eq.~(\ref{equ:component}) and 
of the cross sections for the elastic breakup, the non-elastic breakup, and the deuteron absorption are also shown in Fig.~\ref{fig:Ri-BU}.
As presented in Table.~\ref{table:comparison}, the elastic breakup cross sections $\sigma_{\mathrm{EB}}$ and the neutron removal cross sections $\sigma_{-n}$ predicted by the present calculation are comparable to those calculated under the similar condition in Ref.~\cite{Neoh}.
Note that the target nuclei are $^{93}\mathrm{Zr}$ in the present calculation and $^{90}\mathrm{Zr}$ in Ref.~\cite{Neoh}, respectively, and $\sigma_{-n}$ is defined as follows
\begin{equation}
\label{equ:n-removal}
\sigma_{-n} = \sigma_{\mathrm{EB}} + \sigma_{n-\mathrm{NEB}}.
\end{equation}

It is well illustrated in Fig.~\ref{fig:pd-ef} that the components by the deuteron absorption are dominant at low incident energies but those by the nucleon absorption increase with an increase in incident energy.
As described in Sec.~\ref{sec:100MeV}, these behaviors are mainly related to the energy distributions of absorbed proton $f(E_p)$ and neutron $f(E_n)$
since the energy dependence of each $R_i$ in Fig.~\ref{fig:Ri-BU}~(a) is not strong as that of each component in Fig.~\ref{fig:pd-ef} and $R_d$ is still the largest at 200 MeV/nucleon.

One can notice that the CCONE calculation is in good agreement with the experimental data for $^{102}\mathrm{Pd}$ in Fig.~\ref{fig:pd-118} and for $^{99}\mathrm{Ru}$ in Fig.~\ref{fig:pd-196}.
However, it is found from Fig.~\ref{fig:pd-ef} that CCONE does not necessarily reproduce the energy dependence of the experimental data.

Significant differences are seen between the calculations of DEURACS and CCONE also at energies below 100 MeV/nucleon.
In addition, at the incident energies below 40 MeV/nucleon where the empirical formula by M. Avrigeanu~{\it et al.}~\cite{AvrBU} is available, 
the formula predicts the larger cross sections for the non-elastic breakup and the lower ones for the deuteron absorption, respectively, in comparison with the present calculation.
Thus, it is expected that their framework predicts the lower production cross sections for the residual nuclei shown in Fig.~\ref{fig:pd-ef} in the energy range below 40 MeV/nucleon.
Measurements for $^{107}\mathrm{Pd}+d$ reactions at lower incident energies will be encouraged for further investigation of the role of the deuteron breakup processes.
This is also important from the viewpoint of providing comprehensive cross section data
because in the actual application the primary deuteron beam may be slowed down in the target system.

Better agreement with the experimental data may be obtained by adjusting the model parameters although it is not the object of the present work.
Generally speaking, a series of model parameters are involved in theoretical models describing the pre-equilibrium and compound processes,
and the parameters are often needed to be optimized for providing accurate cross section data.
The three findings described in the end of Sec.~\ref{sec:100MeV} strongly indicate that the optimization should be carried out in a framework considering the breakup processes properly, such as DEURACS.
They also suggest that the optimizations for model parameters relevant to nucleon-induced reactions will be necessary even for deuteron-induced spallation reactions.
The proton-induced reactions on $^{93}\mathrm{Zr}$ and $^{107}\mathrm{Pd}$ were also favorably measured in Refs.~\cite{Kawase,WangPd}. 
The experimental proton-induced data will be useful for accurate prediction of deuteron cross section on the LLFPs.

\renewcommand{\arraystretch}{2.0}
\begin{table}[h]
	\caption{\label{table:comparison} 
			Elastic breakup cross sections $\sigma_{\mathrm{EB}}$ and neutron removal cross sections $\sigma_{-n}$ at the deuteron total kinetic energy of 56 MeV.
		}
	\begin{ruledtabular}
		\begin{tabular}{lcc}
			&  $\sigma_{\mathrm{EB}}$ [mb] & $\sigma_{-n}$ [mb] \\
			\hline
			 Present ($^{93}\mathrm{Zr}$)                      & 181.14 & 610 \\
			Neoh {\it et al.}~\cite{Neoh} ($^{90}\mathrm{Zr}$) & 172.69 & 590 \\
		\end{tabular}
	\end{ruledtabular}
\end{table}

\subsection{Deuteron total reaction cross sections }
\label{sec:d-rea}
\begin{figure}[h]
	\includegraphics[scale=0.80]{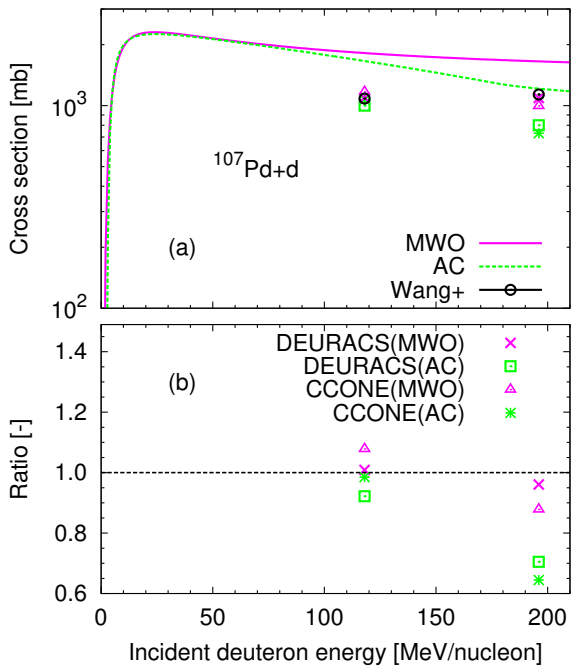}
	\centering
	\caption{\label{fig:wmo-ac} (a) Comparison of the deuteron total reaction cross sections $\sigma_{(d,rea)}$ on $^{107}\mathrm{Pd}$.
		                        The dashed and solid curves represent the $\sigma_{(d,rea)}$ by the MWO formula and the optical model calculation with the AC OPs.                  
		                        The summations of the isotopic production cross sections measured in Ref.~\cite{WangPd} are shown by the open circles. 
	                            The crosses, open squares, open triangles, and asterisks are the calculated values corresponding to the experimental data. 
	                            The crosses and open squares are obtained from the DEURACS calculation using the $\sigma_{(d,rea)}$ by the MWO formula and the optical model calculation with the AC OPs.
                                The open triangles and asterisks are obtained from the CCONE calculation using the two $\sigma_{(d,rea)}$, respectively.
                                (b) The ratio of each calculated value to the experimental data.
                            }
\end{figure}

Finally, we discuss the validity of the MWO formula providing the deuteron total reaction cross sections.
The curves in Fig.~\ref{fig:wmo-ac} represent the $\sigma_{(d,rea)}$ on $^{107}\mathrm{Pd}$ by the MWO formula and the optical model calculation with the AC OPs. 
Both are in good agreement with each other at incident energies up to 100 MeV/nucleon, and in the energy range they have been validated through global comparison with experimental data on other stable nuclei
ranging from $^{12}\mathrm{C}$ to $^{208}\mathrm{Pb}$~\cite{MWO,AnCaiOMP}.
On the other hand, a disagreement is seen above 100 MeV/nucleon and the difference reaches to about a factor of 1.4 at 200 MeV/nucleon.
However, there is no experimental data of $\sigma_{(d,rea)}$ above 100 MeV/nucleon including ones on other nuclei than $^{107}\mathrm{Pd}$,
and we can not judge which value is reliable by direct comparison with experiment.

Hence, we utilize the summation of isotopic production cross sections instead of $\sigma_{(d,rea)}$.
In Fig.~\ref{fig:wmo-ac}~(a), the summations of the isotopic production cross sections measured in Ref.~\cite{WangPd} are shown by the open circles. 
It should be noted that production cross sections of all the residual nuclei have not been measured in Ref.~\cite{WangPd} and thus
the summation of them does not necessarily correspond to $\sigma_{(d,rea)}$.
We obtain the quantities corresponding to the experimental data from the calculations with DEURACS and CCONE using the $\sigma_{(d,rea)}$ by the MWO formula and the optical model calculation with the AC OPs, respectively. 
As described in the end of Sec.~\ref{sec:parameters}, the $\sigma_{(d,rea)}$ are used only for the re-normalization of the calculation results. 

As presented in Fig.~\ref{fig:wmo-ac}, the CCONE calculation using the $\sigma_{(d,rea)}$ with the AC OPs reproduces well the experimental data at 118 MeV/nucleon 
but it is not in good agreement with the data at 196 MeV/nucleon. 
The trend is reversed in the case where the MWO formula are adopted.
Thus, it is difficult to conclude which of the two $\sigma_{(d,rea)}$ is more reliable by the results of CCONE calculation.
The difficulty is caused by the absence of deuteron breakup processes in CCONE.
In the CCONE code, the formation of composite nucleus are overestimated since the elastic breakup is not taken into account.
This leads to increasing productions of residual nuclei.
On the other hand, in the CCONE calculation, composite nuclei are formed only from the deuteron absorption and thus the formations of highly excited nuclei are increased. 
This means that light residual nuclei outside of the range measured in Ref.~\cite{WangPd} are more likely to be produced and then the summation corresponding to the experimental data get small.
The competition between these two effects cause overestimation or underestimation of the experimental data even if the values of $\sigma_{(d,rea)}$ are reasonable.

As for the DEURACS calculation, we can conclude that the experimental data consistently support the $\sigma_{(d,rea)}$ by the MWO formula. 
It is worth mentioning that this is the first validation of the MWO formula above 100 MeV/nucleon through comparison with experimental data.
This validation is meaningful also from the viewpoint of transmutation application since the magnitude of $\sigma_{(d,rea)}$ is critically important to discuss the transmutation efficiency of LLFPs.

In the present study we have performed the CCONE calculation in the framework of DEURACS using the AC OPs as the deuteron OPs, 
and then have re-normalized the calculation results using the $\sigma_{(d,rea)}$ by the MWO formula.
Strictly speaking, this prescription is not a consistent way. 
It is desired to develop global deuteron OPs providing $\sigma_{(d,rea)}$ similar to those by the MWO formula.
This will be necessary as one of our future subjects.

\section{Summary and conclusions}
\label{sec:conclusion}
We have analyzed the isotopic production cross sections of residual nuclei in the $^{93}\mathrm{Zr}+d$ and $^{107}\mathrm{Pd}+d$ spallation reactions at 100--200 MeV/nucleon
and investigated a role of the deuteron breakup processes in these reactions.
The analyses and investigations have been performed using the code system called deuteron-induced reaction analysis code system (DEURACS),
in which the breakup processes are explicitly taken into account.

The cross sections calculated with DEURACS reproduced the experimental data well over a wide mass number range of residual nuclei.
These results have demonstrated the applicability of DEURACS to deuteron-induced spallation reactions at intermediate energies.

In addition, the calculated cross sections were decomposed into the individual components corresponding to the absorption of either neutron or proton in the incident deuteron, or the deuteron itself.
A component-by-component analysis has led to the three findings: 
(i) components of the nucleon absorption have large contributions to the production of residual nuclei near the target nucleus.
(ii) isotopes with the atomic number larger than the target nucleus by one are preferentially produced from the proton absorption.
(iii) reduction of the deuteron absorption leads to good prediction of cross sections for light residual nuclei.
These findings strongly indicate that the consideration of the breakup processes is essential for an accurate
description of deuteron-induced spallation reactions.

Finally, we have confirmed that the deuteron total reaction cross sections $\sigma_{(d,rea)}$ provided by the formula developed by Minomo, Washiyama, and Ogata (MWO) are reliable at 200 MeV/nucleon.
The framework of DEURACS has enabled us to perform this validation.
Further validation of the MWO formula and development of global deuteron optical potentials providing $\sigma_{(d,rea)}$ similar to the formula will be necessary as one of our future subjects.

\begin{acknowledgments}
One of the authors (S.N.) is grateful to H. Wang for the fruitful discussions.
This work was funded by ImPACT Program of Council for Science, Technology and Innovation (Cabinet Office, Government of Japan).
\end{acknowledgments}

\end{document}